\documentclass[aps,prl,amsmath,amssymb, reprint, superscriptaddress,]{revtex4-2}

\usepackage{nicefrac} 
\usepackage{graphicx}
\usepackage{dcolumn}
\usepackage{bm}
\usepackage[mathlines]{lineno}

    \usepackage{lineno}  
    \usepackage{amsmath} 
    \usepackage{etoolbox} 
        
\usepackage{placeins}

\begin{document}

\title{Spontaneously-Induced Dirac Boundary State and Digitization in a Nonlinear Resonator Chain}

\author{Gengming Liu}
\affiliation{Department of Physics, University of Illinois at Urbana-Champaign, Urbana, Illinois 61801, USA.}
\author{Jiho Noh}
\affiliation{Department of Mechanical Science and Engineering, University of Illinois at Urbana-Champaign, Urbana, Illinois 61801, USA.}%
\author{Jianing Zhao}
\affiliation{Department of Mechanical Science and Engineering, University of Illinois at Urbana-Champaign, Urbana, Illinois 61801, USA.}%
\author{Gaurav Bahl}
 \email{bahl@illinois.edu}
\affiliation{Department of Mechanical Science and Engineering, University of Illinois at Urbana-Champaign, Urbana, Illinois 61801, USA.}%

\date{\today}

\begin{abstract}
The low-energy excitations in many condensed matter and metamaterial systems can be well described by the Dirac equation. The mass term associated with these collective excitations, also known as the Dirac mass, can take any value and is directly responsible for determining whether the resultant band structure exhibits a band gap or a Dirac point with linear dispersion. Manipulation of this Dirac mass has inspired new methods of band structure engineering and electron confinement.   
Notably, it has been shown that a massless state necessarily localizes at any domain wall that divides regions with Dirac masses of different signs. These localized states are known as Jackiw-Rebbi-type (JR-type) Dirac boundary modes and their tunability and localization features have valuable technological potential. 
In this study, we experimentally demonstrate that nonlinearity within a 1D Dirac material can result in the spontaneous appearance of a domain boundary for the Dirac mass. 
Our experiments are performed in a dimerized magneto-mechanical metamaterial that allows complete control of both the magnitude and sign of the local material nonlinearity, as well as the sign of the Dirac mass. We find that the massless bound state that emerges at the spontaneously appearing domain boundary acts similarly to a dopant site within an insulator, causing the material to exhibit a dramatic binary switch in its conductivity when driven above an excitation threshold.

\end{abstract}

\maketitle

The Dirac equation and its analogues are frequently encountered in a wide range of condensed matter and metamaterial systems 
\cite{shen_topological_2012, lumer_self-localized_2013, kariyado_manipulation_2015, calado_ballistic_2015, fransson_magnon_2016, banerjee_granular_2016, dubois_observation_2017, chaunsali_demonstrating_2017, li_designing_2018, wang_band_2019,  jiao_experimentally_2021, chaunsali_stability_2021}. A prominent characteristic of these systems is the linear dispersion relation near crossing points between their bulk bands, commonly known as Dirac points, where the collective excitations appear massless \cite{castro_neto_electronic_2009, wehling_dirac_2014}. The degeneracy at any Dirac point is held by material-specific symmetries which, when broken, can split into a pair of states with Dirac masses of opposite sign. \cite{wehling_dirac_2014}. Such regions of opposite Dirac mass have been observed in graphene  \cite{giovannetti_substrate-induced_2007, hunt_massive_2013, ponomarenko_cloning_2013}, topological insulators \cite{bernevig_quantum_2006, konig_quantum_2007, zhang_topological_2009, knez_observation_2014} and various metamaterial systems such as phononic  \cite{pal_edge_2017, daguanno_nonlinear_2019, chen_chiral_2020} and photonic crystals \cite{dreisow_classical_2010, gorlach_photonic_2019, xiao_surface_2014}.

Notably, it has been shown that there necessarily exists a localized state on any boundary between regions with opposite Dirac masses, which was originally discussed in the context of quantum field theory by Jackiw and Rebbi \cite{jackiw_solitons_1976} but has since been shown to occur in classical Dirac materials \cite{semenoff_domain_2008, yao_edge_2009, longhi_photonic_2010, angelakis_probing_2015, tan_photonic_2015, tran_linear_2017, sharabi_topological_2018}. As an intuitive interpretation, the trapped state appears because an intermediate massless (and therefore gapless) transition must occur at the boundary for the Dirac mass to flip the sign, which resembles a metallic boundary between two gapped regions. Unlike in the original Jackiw-Rebbi (JR) theory \cite{jackiw_solitons_1976}, a generic boundary state in a Dirac material is not topologically protected to stay at mid-gap \cite{shtanko_robustness_2018}. We therefore describe these modes as being of JR-type, and this more generic state is expected to exhibit a variety of tunable in-gap dispersion such as flat bands and linear crossings. Even so, the localization on the boundary is unchanged and the JR-type Dirac boundary mode is believed to contribute to long-range edge transport observed in a variety of Dirac materials \cite{knez_observation_2014, Ma_unexpected_2015, Nichele_Edge_2016}.

In this work, we show that a domain boundary of Dirac mass, and hence a localized JR-type state, can spontaneously emerge in a Dirac material with Duffing (third-order) nonlinearity. We experimentally implement this using a magneto-mechanical approach that provides complete control over the sign and magnitude of the nonlinear coefficients \cite{grinberg_magnetostatic_2019}. We first confirm that a designed inversion of the Dirac mass in a linear system leads to the appearance of the JR-type state. Subsequently, we introduce the nonlinearity, and show that a Dirac mass boundary and JR-type state can spontaneously appear in a homogeneous system under strong driving conditions. Interestingly, we find that the induced JR-type state acts similarly to a dopant site in an insulator, resulting in a binary switch in the conductivity through the system as soon as it appears.

\begin{figure*}
    \centering
    \includegraphics[width=\textwidth]{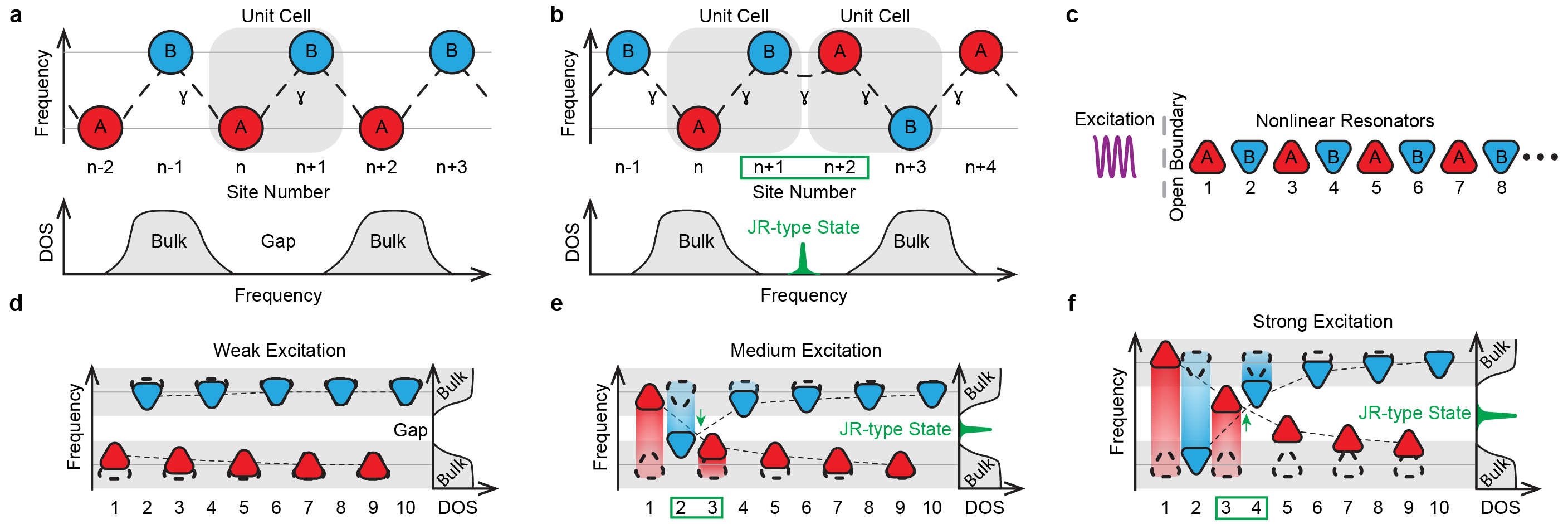}
    \caption{\label{f1}Engineering Dirac boundary mode in linear and nonlinear diatomic arrays. $\mathbf{a)}$ A diatomic chain with staggered resonance frequencies and uniform coupling rate (represented by curved dash lines) has a band gap (lower right) that can be shown through a measurement of the density of state (DOS) below. $\mathbf{b)}$ A JR-type Dirac boundary state arises at the interface where the unit cell resonance frequencies are flipped. The resulting in-gap mode is shown below. $\mathbf{c)}$ For nonlinear arrays, inversion of frequencies can be induced by external excitation from an edge. $\mathbf{d,e,f)}$  Amplitude-dependent frequency detuning in a nonlinear chain at different excitation levels. Dash lines are used as a visual aid to locate the boundary mode.}
\end{figure*}

We start with a brief description of the link between the equations of motion for a diatomic resonator array and the one-dimensional Dirac equation. Consider an infinite one-dimensional dimerized lattice of linear resonators with staggered resonance frequencies and constant nearest-neighbor coupling strength $\gamma$ as shown in Fig.~\ref{f1}(a). To simplify the description for the rest of our paper, we call the sublattices within the unit cell A and B, respectively. It is well known that a mismatch between the natural frequencies of the two resonators (on sublattices A and B) within a unit cell creates two-band insulators \cite{herbold_pulse_2009,maradudin_remarks_1958}. In between these two bands is a forbidden-frequency region (band gap) where no eigenmode exists to host excitation in the structure and thus this diatomic lattice resembles an insulator at such frequencies. 

It has been shown \cite{ruseckas_photonic-band-gap_2011,Tran_optical_2014} that this discrete, classical one-dimensional lattice represented by the coupled-mode equations\begin{equation}  
i\frac{da_n(t)}{dt}+\gamma[a_{n+1}(t)+a_{n-1}(t)]-(-1)^n{\Delta}\cdot a_n(t)=0,
\label{eq1}
\end{equation} can be mapped to the 1D Dirac equation\begin{equation}  
i\partial_t\Psi(\xi,t)=-i\gamma\hat{\sigma}_x\partial_\xi\Psi(\xi,t)+{\Delta}\cdot\hat{\sigma}_z\Psi(\xi,t),
\label{eq2}
\end{equation} after a transformation to the continuous transverse coordinate. Here $a_n$ is the oscillation amplitude at the n$^{th}$ resonator, $2\Delta$ is the resonance frequency mismatch and $\hat{\sigma}_{i}$ are the Pauli matrices. In this analogy, the parameter $\Delta$ is often called the Dirac mass. A more detailed discussion in the context of our system is included in the Supplemental Material \cite{sup}. Based on Eq.~(\ref{eq1}), if we define the first resonator on sublattice site A to be $n=0$, a positive Dirac mass would correspond to a chain with the A-site resonators having a higher resonance frequency than that of B-site ones. At the interface between lattices with Dirac masses of opposite signs [Fig.~\ref{f1}(b)], a localized JR-type state emerges in the band gap. The occurrence of this state has been discussed in previous analytical \cite{tran_linear_2017} and experimental \cite{tan_photonic_2015} studies.

Building upon our discussion in the periodic linear array, we introduce nonlinearities into our model to design a self-induced Dirac mass boundary. Since JR-type modes form at the domain walls separating regions with sign-changing Dirac mass, we can engineer a spontaneously appearing JR-type mode by adding to Eq.~(\ref{eq1}) cubic nonlinearities (described and characterized for our system in Supplemental Material \cite{sup}) that closely resemble the third-order Kerr nonlinearities found in optics. Specifically, we swap the coefficient of the last term $-(-1)^n\Delta$ with\begin{eqnarray}
2\pi\cdot{f}_{n}(a_n)=2\pi[f_n^{(0)}-(-1)^n\cdot sgn(\delta)\cdot\beta a_n^2]\nonumber\\=2\pi\{f_0+(-1)^n[\delta-sgn(\delta)\cdot\beta a_n^2]\},\label{eq3}
\end{eqnarray}where ${f}_n$ is the effective resonance frequency for the $n^{th}$ nonlinear resonator, $\beta>0$ is the third-order nonlinearity coefficient, $f_n^{(0)}\equiv f_0+(-1)^n{\delta}$ are the staggered initial resonance frequencies from the center frequency $f_0$ and $sgn(\cdot)$ is the sign function. In this way, the local Dirac mass is $2\pi[\delta-sgn(\delta)\cdot\beta a_n^2]$ and the stiffening and softening nonlinearities are represented by the positive and negative signs before the nonlinearity coefficient $\beta$ respectively.   
\begin{figure*}[htp!]
    \centering
    \includegraphics[width=\textwidth]{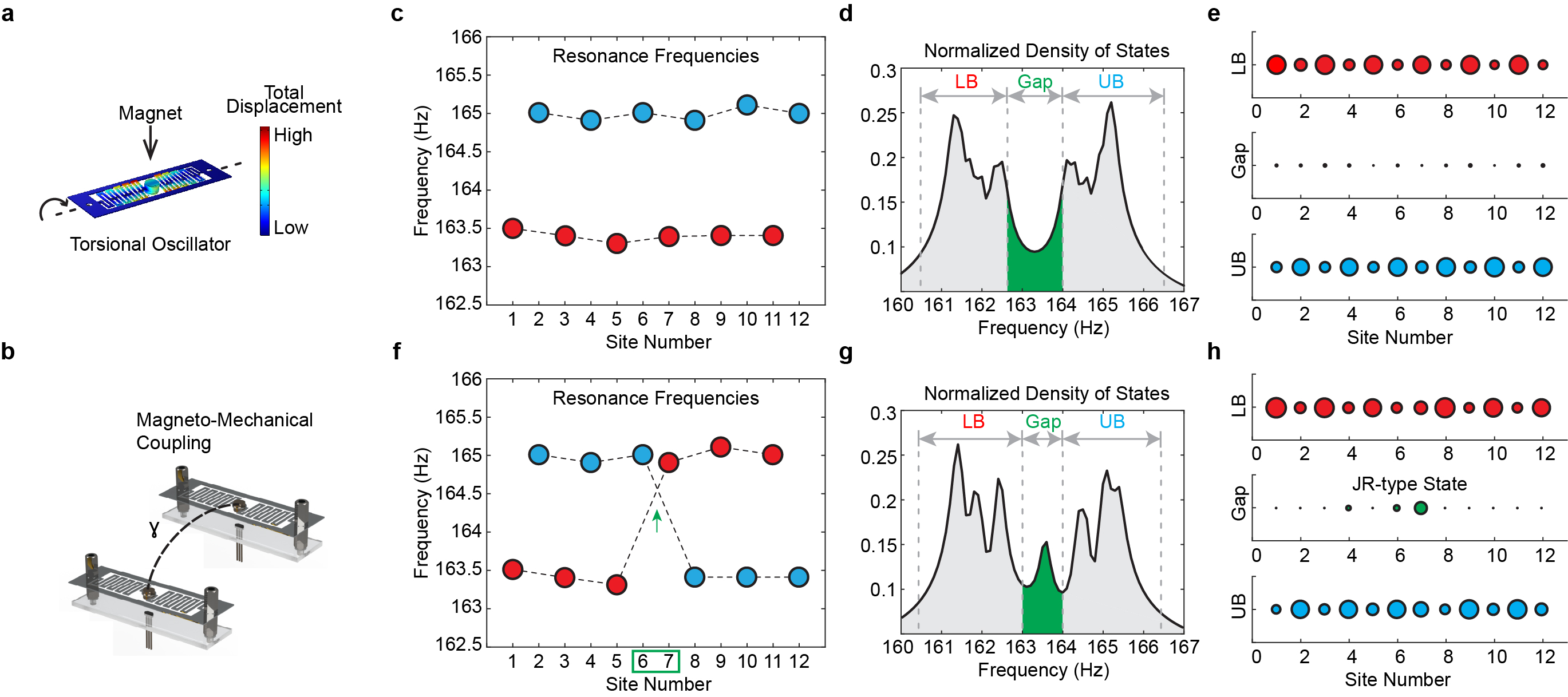}
    \caption{\label{f2}Experimental realization of Dirac edge mode in a linear magneto-mechanical resonator array. $\mathbf{a)}$ The torsional degree of motion superimposed on one resonator model. $\mathbf{b)}$ Magnetic coupling between neighboring resonators. $\mathbf{c,f)}$ Individual resonance frequencies for an array without/with the frequency-inversion interface. $\mathbf{d,g)}$ The resulting density of states shows an empty/occupied band gap (green) between the upper bulk band (UB) and lower bulk band (LB) colored in gray. $\mathbf{e,h)}$ Spatial distributions of the modes are shown in the DOS.}
\end{figure*}

While the nonlinear detuning of a JR mode in purely stiffening and softening photonic metamaterial was already discussed in a previous numerical study \cite{tran_linear_2017}, our experimental effort focuses on inducing a sign flip of the Dirac mass using nonlinear effects and thereby creating a spontaneously appearing in-gap JR-type state.

To induce such a mode in a chain of nonlinear resonators, we tune the initial resonance frequencies into a staggered configuration and impart upon them alternating types of nonlinearities just as described in Eq.~(\ref{eq3}). An excitation is then applied on the left edge of the array [Fig.~\ref{f1}(c)] to induce--at high excitation amplitudes--an interface separating the resonators into two regions with positive and negative Dirac masses through designed nonlinear effects. Specifically, we put softening nonlinearity onto resonators with a positive initial frequency offset $\delta$ and stiffening nonlinearity onto those with a negative offset $-\delta$. In this way, the Dirac mass $2\pi[\delta-sgn(\delta)\cdot\beta a_n^2]$ across this finite chain is linked to the local excitation amplitude $a_n$. In our experiment, the nonlinear array starts with a uniform negative Dirac mass and a clear band gap. At low levels of excitation, even though nonlinearities on resonators near the edge are weakly evoked, no sign change of the Dirac mass is induced, and thus no in-gap mode was observed [Fig.~\ref{f1}(d)]. When the excitation amplitude reaches a certain threshold, the effective resonance frequencies of the resonator pair closest to the array edge become highly detuned and equalize--closing the local band gap. Further increase of the excitation amplitude flips the effective resonance frequencies and changes the sign of the Dirac mass for this segment, creating an interface within our nonlinear array. Based on our previous argument, a JR-type state at an in-gap frequency would emerge at the interface [Fig.~\ref{f1}(e)]. Increasing the excitation level further recruits more resonators pairs to the side with positive Dirac mass and sends the interface deeper into the gapped array [Fig.~\ref{f1}(f)].

Experimental realization of the system requires a type of resonator with tunable resonance frequency as well as stiffening or softening nonlinearities. Magneto-mechanical resonators comprised of an axially magnetized disc magnet mounted on a mechanical torsional spring satisfy these requirements \cite{grinberg_magnetostatic_2019, grinberg_robust_2020, grinberg_trapped_2020}. We produced the torsional springs via water jet cutting of patterns on 0.4-millimeter-thick aluminum plates, such that when displaced from its neutral angular position, the spring provides a restoring torque [Fig.~\ref{f2}(a)]. Axially magnetized N52 neodymium disc magnets (D41-N52, K\&J magnetics, Inc.) are then fixed to the springs to mass-load each resonator while allowing magnetic coupling between them [Fig.~\ref{f2}(b)]. We fine-tuned the natural frequency of every individual resonator by adjusting the mass loading. The coupling strength was controlled by adjusting the distance between neighboring resonators since the magnetic interaction between these resonators decays cubically with distance \cite{grinberg_robust_2020,grinberg_trapped_2020}. This fast decay over distance allows us to consider only the nearest-neighbor coupling in our studies.  
\begin{figure*}
    \centering
    \includegraphics[width=\textwidth]{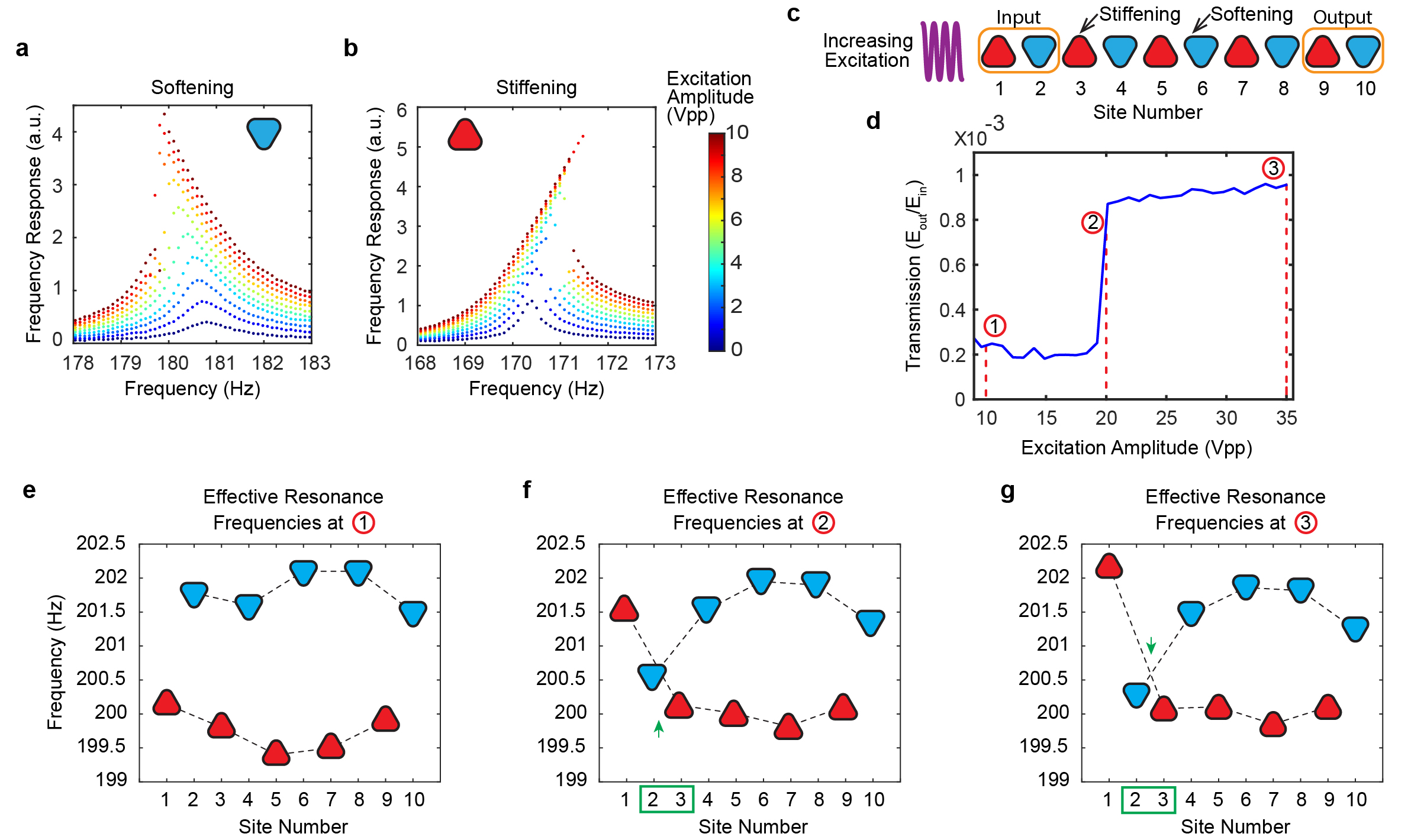}
\caption{\label{f3}Realization of a spontaneously-induced Dirac boundary mode in a nonlinear array. $\mathbf{a,b)}$ Sample frequency response curves for a resonator with spring softening (stiffening) nonlinearity. Note that a different coil system is used in the nonlinear experiment, refer to Supplementary Material \cite{sup}, Fig.~4 for calibrated nonlinear detunings. $\mathbf{c)}$ Configuration for a nonlinear array where energy transmission is defined as output energy over the input (orange boxes). $\mathbf{d)}$ Energy transmission curve as a function of excitation level. $\mathbf{e,f,g)}$ Effective individual resonance frequencies for the nonlinear array at three different excitation levels labeled in the transmission curve.}
\end{figure*}

To study the emergence of the JR-type mode, we first assemble a linear array by placing twelve frequency-tuned resonators into two configurations. For the first configuration, the resonators of high and low resonance frequencies are arranged so that the Dirac mass for the entire array is negative [Fig.~\ref{f2}(c)]. Specifically, resonators on sublattice A and sublattice B have their resonance frequencies tuned to $f_A=163.5$ Hz and $f_B=165.0$ Hz respectively. We keep all resonators at the same height with their torsional axis parallel to one another and set the distance between the centers of the magnets to 2.7 cm for uniform coupling strength.   
A measurement of the mechanical susceptibility is then performed on the entire array to probe its density of states (DOS). We locally excite each resonator along the chain via magnetic torque applied using a coil. The resulting angular displacement is then recorded as a function of time through a Hall sensor placed directly below the resonator. To obtain the DOS, frequency scans across the relevant frequency range are performed for each resonator across the chain and the steady-state responses are recorded in increments of $0.1$ Hz. The DOS of an array with only the negative Dirac mass is then shown in Fig.~\ref{f2}(d) where a clear band gap is observed. We are also able to plot the excitation pattern for the bulk bands as shown in Fig.~\ref{f2}(e). As expected for a dimerized array, the two bulk modes are separated by a band gap. 
For the second configuration, we manually flip the resonance frequencies for half of the chain. At the center of this new chain, the Dirac mass changes from negative to a positive value and we expect to see a JR-type mode localized at this interface [Fig.~\ref{f2}(f)]. Indeed, we are able to identify a clear in-gap state from the DOS plot for the second linear chain [Fig.~\ref{f2}(g)]. At the same time, the spatial distribution of the mode as shown in Fig.~\ref{f2}(h), verifies that this JR-type mode is localized at the lattice interface. Through the comparison between these two different configurations of the
linear array, we confirm the correspondence between an inversion of the effective resonance frequencies for resonators on sublattices A and B and the emergence of a localized JR-type state--a conclusion we will use in the nonlinear experiment next.

In order to introduce nonlinearities into the resonator chain as required by Eq.~(\ref{eq3}), we follow the approach described in a previous study on magnetostatic spring stiffening and softening effects \cite{grinberg_magnetostatic_2019}. Specifically, the nonlinearities can be introduced by placing a fixed neodymium magnet near each resonator to generate a local non-uniform ambient magnetic field. In the point dipole limit, the magneto-mechanical resonator experiences a magnetostatic spring effect in addition to the mechanical restoring force. Depending on the relative location and orientation of the magnetic moments (see Supplemental Material \cite{sup}), the magnetostatic spring effect can produce either softening or stiffening with increasing oscillation amplitude. In Fig.~\ref{f3}(a,b), we show the frequency response of nonlinear softening and stiffening resonators as a function of excitation amplitude. The experimentally measured frequency shifts follow the third-order nonlinearities described in Eq.~(\ref{eq3}). Additionally, we recorded the effective resonance frequencies at different oscillation amplitudes for every resonator used in the nonlinear resonator chain (Supplemental Material \cite{sup}). We later use these amplitude-frequency curves to infer the effective resonance frequencies in the nonlinear experiment.  

We next construct an array of resonators with alternating types of nonlinearities [see Fig.~\ref{f3}(c)] where a JR-type mode spontaneously appears. The softening resonators are placed on the B sublattice and are tuned so that their initial effective resonance frequencies are around $f_B^{(0)}=202.0$ Hz. The stiffening resonators are tuned to $f_A^{(0)}=200.0$ Hz and placed on A sublattice. Ten nonlinear resonators are used to build the resonator array with identical inter-resonator distance (and thus identical coupling rate). 
We magnetically excite the array on the left edge at the mid-gap frequency of $f_{drive}=200.8$ Hz by applying a sinusoidal current through a coil placed near the left-most resonator [Fig.~\ref{f3}(c)]. The excitation voltage applied on the drive coil is linearly proportional to the coil current and the drive torque on the first resonator (Supplemental Material \cite{sup}). At each excitation level, we simultaneously record the steady-state oscillation amplitudes for all resonators and infer their effective resonance frequencies using the prerecorded amplitude-frequency relation (details in Supplemental Material \cite{sup}).

To study the energy transmission through the array, we further define an energy transmission coefficient $T$ with the expression \begin{equation}
T\equiv\frac{E_{out}}{E_{in}}=\frac{a_{9}^2+a_{10}^2}{a_{1}^2+a_2^2},\label{eq4}
\end{equation} where $a_n$ represents the magnitude of oscillation for the $n^{th}$ resonator and $E_{in}$($E_{out}$) is the input (output) energy defined by the sum of the squares of the oscillation amplitudes for the first (last) unit cell. The energy transmission shows a sudden jump around an excitation level of 20 peak-to-peak Voltage (Vpp). As we show next, this jump corresponds to the sudden increase in tunneling via the JR-type mode which acts similarly to an in-gap dopant site in an insulator.    

We can confirm the spontaneous appearance of the JR-type mode by examining the inferred resonance frequencies of the nonlinear chain at a few points of interest [Fig.~\ref{f3}(d)]. For low excitation amplitude, as in Fig.~\ref{f3}(e) for 10 Vpp, the nonlinearities are only weakly invoked and the effective resonance frequencies on A sites and B sites do not cross, making the Dirac mass negative throughout the chain. For 20 Vpp, we see the first crossing of the resonance frequencies between A and B sites [Fig.~\ref{f3}(f)]. From our previous configuration in Fig.~\ref{f2}, we deduce that a JR-type mode should have appeared in the band gap near the left edge of this nonlinear chain, which in turn facilitated the increased transmission. As we further increase the excitation amplitude to 35 Vpp, we observe more frequency tuning but the location of the frequency crossing or the JR-type mode does not change significantly and thus $T$ remains stable. In principle, the mode should migrate deeper into the chain with even higher excitation as more sites are recruited into the flipped configuration. Within our experiment range, we observe a digitization behavior in the transmission where $T$ remains at a low ``zero-state'' until the excitation amplitude exceeds the threshold of 20 Vpp where $T$ suddenly jumps up to (and maintains) a higher level resembling a ``one-state''. This digitization behavior not only is helpful in showing the emergence of an in gap Dirac boundary state in our nonlinear resonator array but also readily offers an  example of the application for the JR-type Dirac boundary mode in one dimensional nonlinear metamaterials.      

The ability to control both the magnitude as well as the sign of the excitation mass in Dirac materials has been sought after for its potential to engineer boundary states \cite{giovannetti_substrate-induced_2007, semenoff_domain_2008}. In this paper, we demonstrate a resonator array where local nonlinearities can be laid out in a way such that a sign-changing Dirac mass boundary spontaneously emerges at sufficiently high levels of drive amplitude and a JR-type boundary mode appears. We further show a correspondence between the emergence of this mode and a sudden improvement of in-gap energy transmission. Our work is generalizable to any other systems with nonlinearity. For future studies, we are eager to see similar spontaneously appearing Dirac boundary states being developed in higher dimensions where their robustness to disorder along the Dirac mass boundary and tunable mode dispersion \cite{shtanko_robustness_2018} can be selectively turned on to dynamically modify material properties.

\begin{acknowledgments}
We acknowledge funding support from the U.S. National Science Foundation (NSF) Emerging Frontiers in Research and Innovation program (EFRI) and Office of Naval Research (ONR) Director of Research Early Career Grant (grant N00014-17-1-2209).
The authors additionally thank Dr. Inbar Grinberg, Prof. Taylor Hughes, and Dr. Xiao-Qi Sun for their valuable insights and discussions.
\end{acknowledgments}


%

\end{document}


\title{Supplemental Materials: Spontaneously-Induced Dirac Boundary State and Digitization in a Nonlinear Resonator Chain}

\author{Gengming Liu}
\affiliation{Department of Physics, University of Illinois at Urbana-Champaign, Urbana, Illinois 61801, USA.}
\author{Jiho Noh}
\affiliation{Department of Mechanical Science and Engineering, University of Illinois at Urbana-Champaign, Urbana, Illinois 61801, USA.}
\author{Jianing Zhao}
\affiliation{Department of Mechanical Science and Engineering, University of Illinois at Urbana-Champaign, Urbana, Illinois 61801, USA.}
\author{Gaurav Bahl}
 \email{bahl@illinois.edu}
\affiliation{Department of Mechanical Science and Engineering, University of Illinois at Urbana-Champaign, Urbana, Illinois 61801, USA.}

\date{\today}

\maketitle

\setcounter{equation}{0}
\setcounter{figure}{0}
\setcounter{table}{0}
\setcounter{page}{1}
\makeatletter
\renewcommand{\theequation}{S\arabic{equation}}
\renewcommand{\thefigure}{S\arabic{figure}}
\renewcommand{\bibnumfmt}[1]{[S#1]}
\renewcommand{\citenumfont}[1]{S#1}

\section{\label{EOM}Equations of motion and their link to the 1D Dirac Equation}

\begin{figure}[hb]
     \centering
     \includegraphics[width=.5\linewidth]{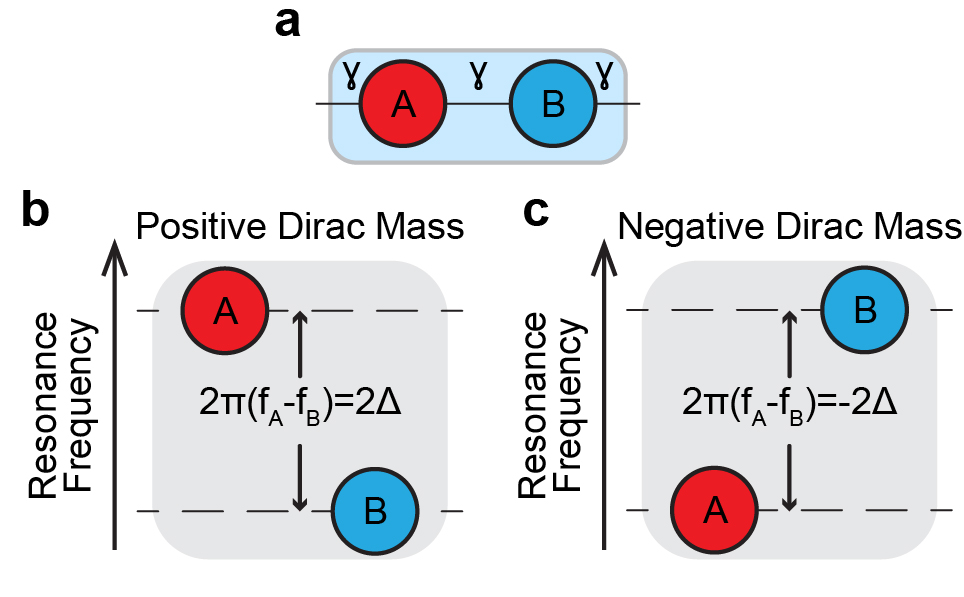}
     \caption{Toy models for dimerized periodic resonator chain and configurations for positive/negative Dirac mass as used in the main text.}\label{s1}
\end{figure}

\begin{figure}[tb]
     \centering
     \includegraphics[width=.35\linewidth]{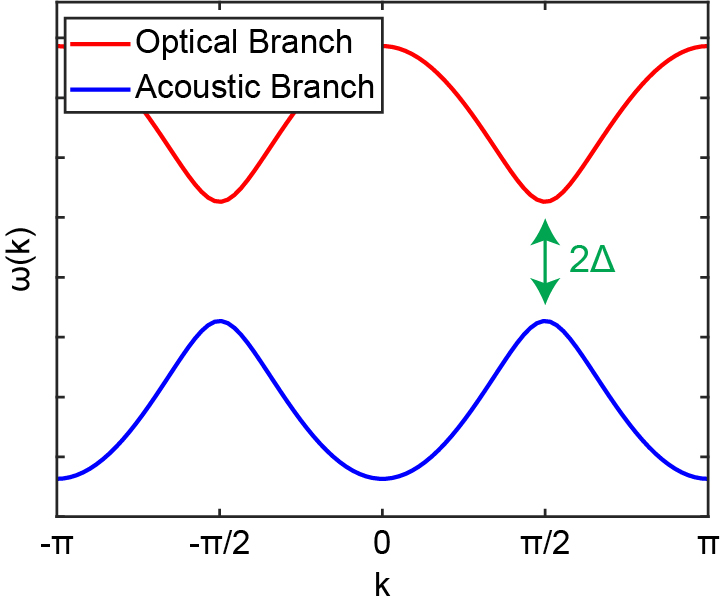}
     \caption{Dispersion of the linear array shown in Fig.~1(a) of the main text \cite{main}. The gap size is labeled with green arrows.}\label{s2}
\end{figure}

Consider a periodic array having two resonators per unit cell [Fig.~\ref{s1}(a)]. We define the displacement of each resonator $x_n(t)$ where $n$ represents the resonator index. From here, we can describe our system using the coupled-mode equations 

\begin{equation}
      i\frac{d x_n(t)}{dt}+\gamma[x_{n+1}(t)+x_{n-1}(t)]-(-1)^n\Delta\cdot x_n(t)=0,
\end{equation}
where $\gamma$ is the coupling strength, $2\Delta$ is the resonance frequency mismatch between the resonators within each unit cell. The periodic array supported by this system of equations has two bands separated by a band gap of size $2\Delta$ (Fig.~\ref{s2}) defined by the dispersion curves $\omega_{\pm}(k)=\pm\sqrt{\Delta^2+4\gamma^2\cos^2(k)}$, assuming unit lattice constant.

Near the edge of the first Brillouin zone, for non-zeros $\Delta$, the dispersion curves form two opposite hyperbolas--mimicking the typical energy-momentum relation of a freely moving relativistic massive particle \cite{dreisow_classical_2010}. In this way, phonons moving in the lattice with wave number $k\approx\pi/2$ simulate the temporal dynamics of the relativistic Dirac equation.

If we transform the coupled-mode equations by setting
\begin{equation}
  \begin{cases}
    \begin{aligned}
& x_{2n}(t)=(-1)^n\psi_1(n,t),\\
& x_{2n-1}(t)=-i(-1)^n\psi_2(n,t),
    \end{aligned}
  \end{cases}
\end{equation} and switching to the continuous transverse coordinate $n\to\xi$, the two-component spinor wave function describing displacements in an unit cell $\psi(\xi,t)=(\psi_1,\psi_2)^T$ satisfies the one-dimensional Dirac equation for a freely moving particle of mass $m$\begin{equation}
i\frac{\partial\psi}{\partial t}=-i\gamma\hat\sigma_x\frac{\partial\psi}{\partial\xi}+\Delta\hat\sigma_z\psi,
\end{equation} where the formal change is made
$\gamma\to c$, $\Delta\to\frac{mc^2}{\hbar}$.

We therefore recognize the term $\Delta$ as the Dirac mass for a periodic array which could take on either positive or negative values [Fig.~\ref{s1}(b,c)]. Furthermore, it has been shown that at the interface between two such lattices with Dirac mass of opposite signs ($\Delta_1=-\Delta_2$), a Jackiw-Rebbi-type edge state will emerge \cite{tran_linear_2017}.

In our work, we tune the Dirac mass terms along an array by adding third-order nonlinearities to our resonators such that a JR-type mode would spontaneously emerge as the injected excitation level increases above a certain threshold [See Main Text \cite{main}, Fig.~3(d-g)]. In the next section, we explain how we control the nonlinearities of our resonators.\\

\begin{figure}[tb]
    \centering
    \includegraphics[width=0.9\linewidth]{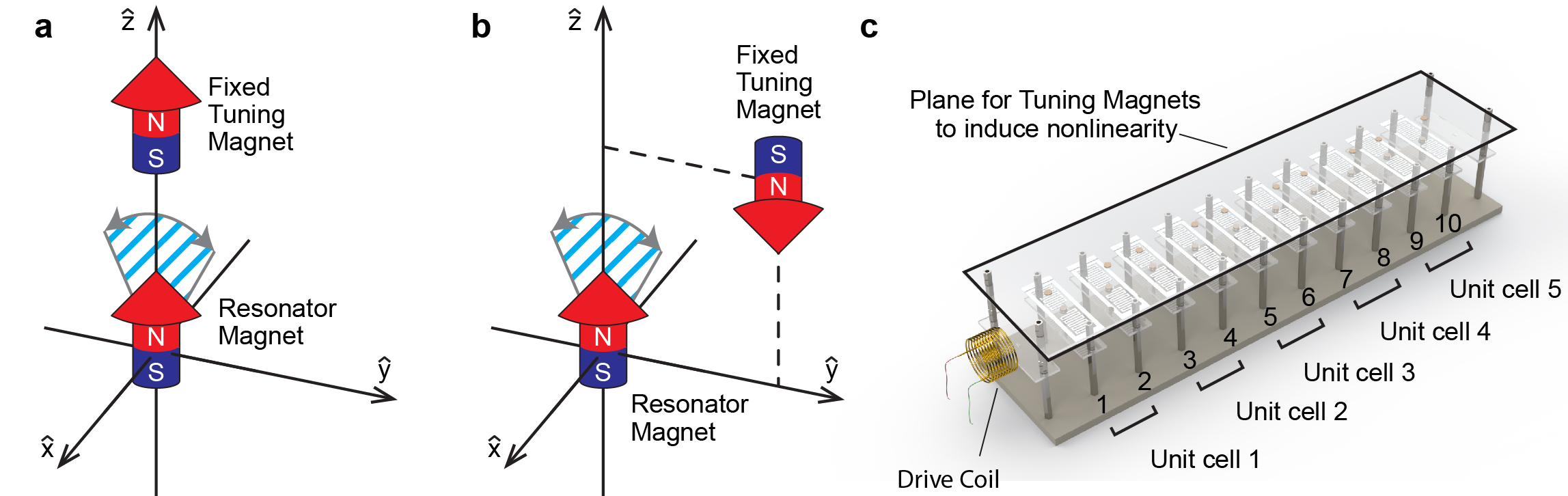}
\caption{\label{s3}Spatial configurations to create cubic nonlinearities. $\mathbf{a,b)}$ Position and orientation of fixed tuning magnets relative to the resonator magnet in the softening and stiffening configurations, respectively. Note, all magnets are aligned on the $\hat x$ direction and the resonator magnets rotates along its $\hat z$ axis which cartoon depicted in exaggeration. $\mathbf{c)}$ Cartoon of the array used in nonlinear experiments.}
\end{figure}

\section{\label{nl}Resonator Nonlinearity Control}

Magnetically sensitive mechanical oscillators in a nonuniform magnetic field can experience a magnetostatic spring effect in addition to the usual mechanical restoring force. Such effect has been studied in detail \cite{grinberg_magnetostatic_2019} and was found to be capable of creating magneto-mechanical resonators with softening as well as stiffening nonlinearities depending on the relative spatial relationship of the magneto-mechanical resonator with respect to the local magnetic field.

We show in Fig.~\ref{s3}(a) and Fig.~\ref{s3}(b) two configurations where the magnetic field from a fixed disc magnet gives the magneto-mechanical resonators softening and stiffening nonlinearities respectively (example data shown in Fig.~3 of Main Text \cite{main}). Then, by adding tuning magnets, we are able to construct a compact resonator array with alternating types of nonlinearities as shown in Fig.~\ref{s3}(c). A drive coil is then used to excite the array from its left edge.

\section{\label{fs}Inferring Resonator Frequencies in a Nonlinear Experiment}

\begin{figure*}[!tbph]
    \includegraphics[width=0.90\textwidth]{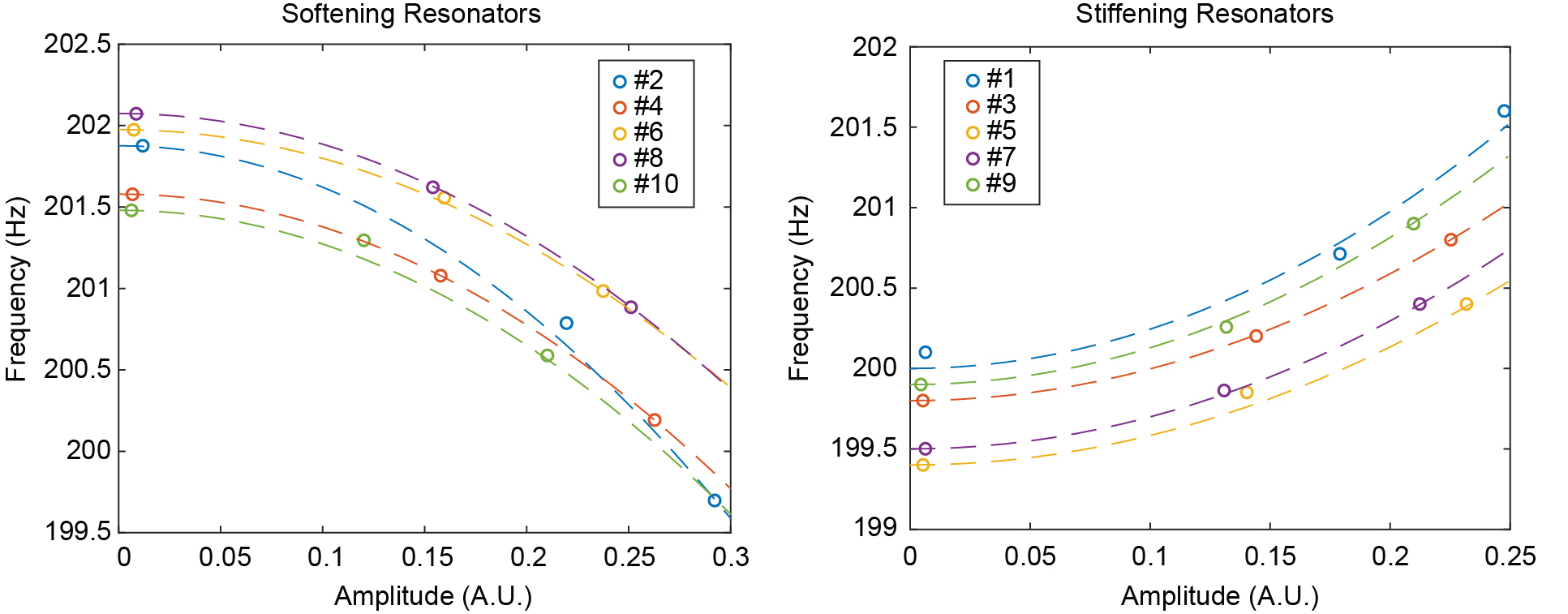}
    \caption{Individually measured nonlinear detuning for all softening and stiffening resonators from the nonlinear experiment. Detuning and frequency response are measured in a similar fashion as shown in Main Text \cite{main}, Fig.~3(a and b) at three excitation amplitudes for analysis, respectively.}\label{s4}
\end{figure*}

Prior to assembling the nonlinear array, we calibrate each nonlinear resonator and the accompanying Hall sensor using three drive amplitudes so that the nonlinear frequency detuning can be measured against the oscillation amplitudes (shown in Fig.~\ref{s4} for softening and stiffening resonators). Fitting of these amplitude-frequency nonlinearities is then done using Main Text \cite{main}, Eq.~(3) for each resonator.

In the experiment, we measure the displacements of all resonators and acquire the steady-state oscillation amplitudes for them at each excitation level. These measured amplitudes are then used to back-extract the effective individual resonance frequencies shown in the Main Text \cite{main}, Fig.~3(e-g). 

\FloatBarrier
%